\documentclass[aps,prc,twocolumn,10pt, superscriptaddress, showpacs, floatfix]{revtex4-1}

\usepackage{amssymb,epsfig}

\newcommand{\be}{\begin{equation}}
\newcommand{\ee}{\end{equation}}
\newcommand{\bea}{\begin{eqnarray}}
\newcommand{\eea}{\end{eqnarray}}
\begin{document}

\title{Many-body correlations in nuclear superfluidity}
\author{Elena Litvinova}
\affiliation{Department of Physics, Western Michigan University, Kalamazoo, MI 49008, USA}
\affiliation{National Superconducting Cyclotron Laboratory, Michigan State University, East Lansing, MI 48824, USA}
\affiliation{GANIL, CEA/DRF-CNRS/IN2P3, F-14076 Caen, France}
\author{Peter Schuck}
\affiliation{Institut de Physique Nucl\'eaire, IN2P3-CNRS, Universit\'e Paris-Sud, F-91406 Orsay Cedex, France}
\affiliation{Universit\'e Grenoble Alpes, CNRS, LPMMC, 38000 Grenoble, France}


\date{\today}

\begin{abstract}
The two-fermion two-time correlation function in the pairing channel is discussed within the equation of motion framework.  Starting from the bare two-fermion interaction, we derive the equation of motion for the two-fermion pair propagator in a strongly-correlated medium. The resulting equation is of the Dyson type
with the kernel having static and one-frequency dependent components and, thus, can be regarded as a Dyson Bethe-Salpeter equation (Dyson-BSE). The many-body hierarchy generated by the dynamical interaction kernel is truncated on the level of two-body correlation functions, thus neglecting the explicit  three-body and higher-rank correlations. The truncation is performed via a cluster expansion of the intermediate three-particle-one-hole correlation function irreducible in the particle-particle channel, that leads to the coupling between single fermions and emergent bosonic quasibound states (phonons). The latter couplings are, thus, derived in terms of the exact mapping of the in-medium two-fermion correlation functions onto the domain of phonons without introducing new parameters. 
 The approach is applied to calculations of the pairing gaps in medium-mass nuclear systems, that include calcium, nickel and tin isotopic chains.   

\end{abstract}

\maketitle

\section{Introduction} 

Theoretical description of strongly-interacting many-body systems remains one of the most difficult areas of physical sciences and, despite the many years of effort, still requires 
more elaborate modeling.  An accurate treatment of many-body correlations is the key to unraveling
 the mechanisms of emergent phenomena in strongly-coupled systems at various scales of physics,
 however, it is very difficult in the non-perturbative regimes.

Atomic nuclei are among the systems, where not only the many-body correlations are extremely difficult to treat in an accurate and systematic way, but even the underlying nucleon-nucleon forces are not known precisely. Being rooted in quantum chromodynamics (QCD) on the fundamental level, the nucleon-nucleon interaction still can not be consistently derived from QCD in the form of potentials. The latter can be, instead, modeled by the meson-nucleon dynamics and parametrized by scattering data \cite{MachleidtHolindeElster1987,Machleidt2016}. 
However, the use of such potentials in the standard many-body frameworks does not yet lead to an accurate description of nuclear phenomena. Thus, both the nucleon-nucleon interactions and the strongly-coupled many-body models require further refinement.  

One of the most interesting problems is the understanding and predictive description of  phenomena related to nuclear superfluidity.
It was noticed shortly after the appearance of the Bardeen-Cooper-Schrieffer (BCS) theory of superconductivity \cite{Bardeen1957} that atomic nuclei behave in some respects similarly to  superconducting metals \cite{Bohr1958}. Indeed, the reduction of nuclear moments of inertia, compared to the case of rigid rotation, the odd-even mass differences, low-lying vibrational states, nuclear shapes and level densities can only be reproduced under the assumption of the presence of an interaction acting
between particles with equal and opposite momenta (superfluid pairing, or pairing). Thus, over decades the BCS and the more general Bogoliubov's theory are widely used for the
description of open-shell nuclei \cite{Bogolyubov1958,Soloviev1986,RingSchuck1980}. 

It has become clear quite early that the underlying mechanism of nuclear pairing can be more complex than it is implied in the BCS and Bogoliubov's approaches. 
For instance, the coupling between the single-particle and emergent 
degrees of freedom (phonons), which plays a significant role in the formation of the nuclear ground and excited states \cite{BelyaevZelevinsky1966,BelyaevZelevinsky1966a,BohrMottelson1969,BohrMottelson1975,Broglia1976,BortignonBrogliaBesEtAl1977,BertschBortignonBroglia1983,Soloviev1992}, may also affect nuclear pairing. This idea was investigated in various phenomenological frameworks \cite{VanderSluys1993,Avdeenkov1999,Avdeenkov1999a,BarrancoBrogliaGoriEtAl1999,BarrancoBortignonBrogliaEtAl2005,IdiniPotelBarrancoEtAl2015} that concluded, in particular, that coupling between nucleons and collective surface vibrations (particle-vibration coupling, or PVC) can be responsible for a large fraction of the nuclear pairing. The PVC effects are widely known to be of prime importance in electronic condensed matter systems, where they can even reverse the sign of the repulsive Coulomb interaction to give rise to superconductivity \cite{50BCSorig,50BCS}.


The common deficiency of the state-of-the-art approaches to nuclear structure, which may also affect the current understanding of  nuclear superfluidity, is that they tend to combining different techniques for approaching the static and dynamical parts of the in-medium interaction.
This, however, may lead to uncontrollable inconsistencies, double counting and missing effects, especially those of collective character. 
Indeed, the exact equations of motion (EOM) for the many-body fermionic correlation functions, which are known across the many areas of quantum physics from condensed matter to  quantum chemistry \cite{Tiago2008,Martinez2010,Sangalli2011,Olevano2018}, show that the static and dynamical kernels of these equations are derivable within the EOM framework on equal grounds from the same underlying bare interaction.
A consistent treatment of both kernels in a unified framework is, therefore, necessary for reproducing the collective emergent phenomena from first principles. This has been justified to be a feasible while yet a highly accurate approach, if the infinite EOM hierarchies 
are truncated by cluster expansions of the dynamical kernels in terms of the many-body correlation functions corresponding to the relevant degrees of freedom \cite{SchuckTohyama2016,Olevano2018}. For instance, in the low and intermediate-energy regimes of nuclear physics truncations 
on the level of two-body or three-body correlation functions should be sufficient for a highly accurate approach applicable to a wide range of nuclear phenomena. 

The model-independent EOM method \cite{Schuck1976,AdachiSchuck1989,Danielewicz1994,DukelskyRoepkeSchuck1998,SchuckTohyama2016}  was shown to produce a hierarchy of approximations to the dynamical kernels of the equations for one-fermion and two-time two-fermion propagators. As we discussed in our recent study of Ref. \cite{LitvinovaSchuck2019}, the non-perturbative versions of those kernels, approximated by cluster expansions in terms of the single-particle, particle-hole and particle-particle correlation functions, can be mapped to the kernels of the phenomenological nuclear field theories (NFT) \cite{BohrMottelson1969,BohrMottelson1975,Broglia1976,BortignonBrogliaBesEtAl1977,BertschBortignonBroglia1983,KamerdzhievTertychnyiTselyaev1997,LitvinovaRingTselyaev2007} and quasiparticle-phonon models (QPM) \cite{Soloviev1992,Ponomarev2001,SavranBabilonBergEtAl2006,Andreozzi2008}. 
This mapping provides an understanding of the emergent collective phenomena and a microscopic foundation for the effective Hamiltonians used in these models, connecting the bare nucleon-nucleon interaction and its modification in the strongly-coupled medium. Moreover, such insights allow for considerable extensions of this type of theories to more complex correlations, which are necessary for achieving spectroscopic accuracy in the description of atomic nuclei in a wide energy range.

In Ref. \cite{LitvinovaSchuck2019} such an extension was presented and implemented numerically on the base of the covariant density functional theory, thus advancing the previously developed relativistic version of the NFT \cite{LitvinovaRing2006,LitvinovaRingTselyaev2008,LitvinovaRingTselyaev2010,Tselyaev2013,AfanasjevLitvinova2015,RobinLitvinova2016}.  Although the latter demonstrated a noticeable progress in implementing the PVC models within the covariant self-consistent framework and described satisfactorily some low-energy nuclear phenomena \cite{EndresLitvinovaSavranEtAl2010,PoltoratskaFearickKrumbholzEtAl2014,MassarczykSchwengnerDoenauEtAl2012,LanzaVitturiLitvinovaEtAl2014,Oezel-TashenovEndersLenskeEtAl2014}, it was still lacking the spectroscopic accuracy because of the absence of more complex correlations than those included in the conventional NFT. In Ref. \cite{LitvinovaSchuck2019} we have shown, in particular, that the higher-order correlations beyond the two-quasiparticle-plus-phonon (2q$\otimes$phonon) ones, for instance, 2q$\otimes$2phonon configurations can introduce some further improvements in the description of the nuclear spectra at both low and high energies.  
Other types of correlations, which are rarely addressed in the literature on the NFT and other PVC models, such as the PVC-induced ground state correlations \cite{Kamerdzhiev1991,Tselyaev1989,KamerdzhievTertychnyiTselyaev1997,Robin2019} and the coupling to charge-exchange phonons \cite{Litvinova2016,RobinLitvinova2018} were also shown to be important for spectroscopically-accurate theories. Recent finite-temperature extensions can be found in Refs. \cite{LitvinovaWibowo2018,WibowoLitvinova2019,WibowoLitvinova2019a,LitvinovaRobinWibowo2020}.

In this article we continue to elaborate on the EOM method for fermionic correlation functions and its connections to the phenomenological NFT's. While Ref. \cite{LitvinovaSchuck2019} was focused on the  single-particle and two-time particle-hole fermionic  propagators, here we discuss the two-time two-fermion propagator and the associated pairing gap equation. 
Some of the closely related ideas on the theory of correlated fermion pairs and ab-initio particle-vibration coupling approach were discussed recently in Ref. \cite{Schuck2019}. The formalism starts along the lines of Refs. \cite{DukelskyRoepkeSchuck1998,Schuck2019} and then advances to non-perturbative approximations for the dynamical interaction kernel. 
In the theoretical sections we discuss fermionic Hamiltonians with unspecified interactions, while the equations of motion are confined by the two-body interactions. The theory can be naturally extended to multiparticle forces and bosonic degrees of freedom. 


\section{Fermionic propagators in a correlated medium}
\label{Propagators}
The formalism of correlation functions, such as the Green functions, or propagators, is one of the most convenient and powerful ones in the description of phenomena that occur in strongly-coupled media.
The propagators are directly related to observed excitation spectra and ground state properties of the many-body systems.

The single-fermion propagator is commonly defined as: 
\be
G(1,1') \equiv G_{11'}(t-t') = -i\langle T \psi(1){\psi^{\dagger}}(1') \rangle,
\label{spgf}
\ee
where $T$ is the operator of the chronological ordering and $\psi(1),{\psi^{\dagger}}(1)$ are the one-fermion fields in the Heisenberg picture:
\be
\psi(1) = e^{iHt_1}\psi_1e^{-iHt_1}, \ \ \ \ \ \ {\psi^{\dagger}}(1) = e^{iHt_1}{\psi^{\dagger}}_1e^{-iHt_1},
\ee
while the subscript '1' stands for the full set of the single-particle quantum numbers in a given representation. In the present work the fermionic degrees of freedom are associated with nucleons which compose a many-nucleon system.
The averaging in Eq. (\ref{spgf}) and in the following is performed over the formally exact correlated ground state while the time evolution is determined by the many-body Hamiltonian
\be
H = H^{(1)} + V^{(2)} + W^{(3)} + ...
\label{Hamiltonian}
\ee
Here the operator $H^{(1)}$ is the one-body contribution to the Hamiltonian:
\be
H^{(1)} = \sum_{12} t_{12} \psi^{\dag}_1\psi_2 + \sum_{12}v^{(MF)}_{12}\psi^{\dag}_1\psi_2 \equiv \sum_{12}h_{12}\psi^{\dag}_1\psi_2
\label{Hamiltonian1}
\ee
with the matrix elements $h_{12}$ which, in general, combine the kinetic energy $t$ and the mean-field $v^{(MF)}$ part  of the interaction. The operator $V^{(2)}$ describes the two-body sector associated with the two-fermion interaction
\be
V^{(2)} = \frac{1}{4}\sum\limits_{1234}{\bar v}_{1234}{\psi^{\dagger}}_1{\psi^{\dagger}}_2\psi_4\psi_3,
\label{Hamiltonian2}
\ee
and the operator $W^{(3)}$ generates the three-body forces
\be
W^{(3)} = \frac{1}{36}\sum\limits_{123456}{\bar w}_{123456}{\psi^{\dagger}}_1{\psi^{\dagger}}_2{\psi^{\dagger}}_3\psi_6\psi_5\psi_4
\ee
with the antisymmetrized matrix elements ${\bar v}_{1234}$ and ${\bar w}_{123456}$, respectively. The ellipsis in Eq. (\ref{Hamiltonian}) stands for further multiparticle forces which can be, in principle, included in the theory. We will make an explicit derivation of the equations of motion assuming that the Hamiltonian is confined by the two-body interaction, however, the theory can be naturally extended to multiparticle forces. 


We will work in the 
basis, which diagonalizes the one-body (also named single-particle or mean-field) part of the Hamiltonian (\ref{Hamiltonian1}): $h_{12} =  \delta_{12}\varepsilon_1$. We will see, however,  that on the way to the final equations of motion this basis should be redefined as soon as the one-body part of the Hamiltonian absorbs additional contributions from the two-body sector. 

The fermionic field operators satisfy the anticommutation relations:
\bea
[\psi_1,{\psi^{\dagger}}_{1'}]_+ \equiv \psi_1{\psi^{\dagger}}_{1'}  +  {\psi^{\dagger}}_{1'}\psi_1 = \delta_{11'}, \nonumber \\
\left[ \psi_1,{\psi}_{1'} \right]_{+}  = \left[ {\psi^{\dagger}}_1,{\psi^{\dagger}}_{1'}\right]_+ = 0.
\label{anticomm}
\eea
The Fourier transform of the single-particle propagator (\ref{spgf}), which depends explicitly on the time difference $\tau = t-t'$, is known as  the spectral (Lehmann) representation:
\be
G_{11'}(\varepsilon) = \sum\limits_{n}\frac{\eta^{n}_{1}\eta^{n\ast}_{1'}}{\varepsilon - \varepsilon_{n}^{+} +i\delta} 
+ \sum\limits_{m}\frac{\chi^{m}_{1}\chi^{ m\ast}_{1'}}{\varepsilon + \varepsilon_{m}^{-}-i\delta}
\label{spgfspec}
\ee
with the poles $\varepsilon_{n}^{+} = E^{(N+1)}_{n} - E^{(N)}_0$ and $-\varepsilon_{m}^{-} = -(E^{(N-1)}_{m} - E^{(N)}_0)$ at the energies of the states in $(N+1)$- and $(N-1)$-particle systems related to the ground state of the initial (reference) $N$-particle system. The residues of Eq. (\ref{spgfspec}) are, in turn, composed of matrix elements of the field operators between the ground state $|0^{(N)}\rangle$ of the $N$-particle system and the states $|n^{(N+1)} \rangle$ and $|m^{(N-1)} \rangle$ of the $(N+1)$- and $(N-1)$-particle systems, respectively:
\be
\eta^{n}_{1} = \langle 0^{(N)}|\psi_1|n^{(N+1)} \rangle , \ \ \ \ \ \ \ \  \chi^{m}_{1} = \langle m^{(N-1)}|\psi_1|0^{(N)} \rangle .
\ee
These matrix elements represent the weights of the given single-particle (single-hole) configuration on top of the ground state $|0^{(N)}\rangle$ in the formally exact $n$-th ($m$-th) state of the systems with $(N+1)$ and $(N-1)$ particles. The residues correspond to the observed occupation probabilities of the corresponding states and related to the spectroscopic factors.

In analogy to Eq. (\ref{spgf}) the two-fermion, three-fermion and, in general, $n$-fermion propagators are defined as follows:
\bea
&G&(12,1'2') = (-i)^2\langle T \psi(1)\psi(2){\psi^{\dagger}}(2'){\psi^{\dagger}}(1')\rangle, 
\label{ppGF} \\
&G&(123,1'2'3') = (-i)^3\langle T \psi(1)\psi(2)\psi(3){\psi^{\dagger}}(3'){\psi^{\dagger}}(2'){\psi^{\dagger}}(1')\rangle, \nonumber \\
&G&(12...n,1'2'...n') = \nonumber \\ &=&(-i)^n\langle T \psi(1)\psi(2)...\psi(n){\psi^{\dagger}}(n')...{\psi^{\dagger}}(2'){\psi^{\dagger}}(1')\rangle.
\label{mbgfs}
\eea
In this work we will focus on the two-time two-fermion Green function (\ref{ppGF}) with  $t_1 = t_2 = t, t_{1'} = t_{2'} = t'$, which depends on the single time difference $t-t'$.
In this case, with the help of Eqs. (\ref{anticomm}), Eq. (\ref{ppGF}) can be transformed to the energy (frequency) domain as:
\be
iG_{12,1'2'}(\omega) =  \sum\limits_{\mu} \frac{\alpha^{\mu}_{21}\alpha^{\mu\ast}_{2'1'}}{\omega - \omega_{\mu}^{(++)}+i\delta} - \sum\limits_{\varkappa} \frac{\beta^{\varkappa\ast}_{12}\beta^{\varkappa}_{1'2'}}{\omega + \omega_{\varkappa}^{(--)}-i\delta},
\label{resppp}
\ee
where the poles $\omega_{\mu}^{(++)} = E_{\mu}^{(N+2)} - E_0^{(N)}$ and  $\omega_{\varkappa}^{(--)} = E_{\varkappa}^{(N-2)} - E_0^{(N)}$ are the formally exact states of the systems with $(N+2)$  and $(N-2)$ particles, respectively,
and the residues are the products of the matrix elements:
\be
\alpha_{12}^{\mu} = \langle 0^{(N)} | \psi_2\psi_1|\mu^{(N+2)} \rangle , \ \ \ \ \ \ \beta_{12}^{\varkappa} = \langle 0^{(N)} | \psi^{\dagger}_2\psi^{\dagger}_1|\varkappa^{(N-2)} \rangle .
\ee
As they connect the states of the $(N+2)$-  and $(N-2)$-particle systems to the ground state of the initial $N$-particle system,  
the two-body propagator of Eq. (\ref{resppp}) describes the response to the probes with pair transfer (addition and removal of two fermions, respectively).
For further analysis it is convenient to include the phase factor "i" into the two-body propagator, so that here we start to use the modified definition:
\be
G(12,1'2') = -i\langle T \psi(1)\psi(2){\psi^{\dagger}}(2'){\psi^{\dagger}}(1')\rangle, 
\label{ppGFmod} \\
\ee
i.e. replace $iG(12,1'2') \to G(12,1'2')$.

Earlier in Ref. \cite{LitvinovaSchuck2019} we have considered the two-time two-fermion propagator in the particle-hole channel. It was shown, in particular, that an accurate description of this response function leads to an EOM with the dynamical kernel, where the particle-hole ($ph$) and particle-particle ($pp$) channels are coupled.  Similarly, we will see below that the EOM for the $pp$-channel, which describes the propagator of Eqs. (\ref{ppGF},\ref{resppp}), will require the knowledge about the particle-hole response function:
\be
R(12,1'2') =  -i\langle T\psi^{\dagger}(1)\psi(2)\psi^{\dagger}(2')\psi(1')\rangle,
\label{phresp}
\ee
which also depends on two times as  $t_1 = t_2 = t, t_{1'} = t_{2'} = t'$ and whose spectral image, or Fourier transform, reads:
\be
R_{12,1'2'}(\omega) = \sum\limits_{\nu>0}\Bigl[ \frac{\rho^{\nu}_{21}\rho^{\nu\ast}_{2'1'}}{\omega - \omega_{\nu}+i\delta} -  \frac{\rho^{\nu\ast}_{12}\rho^{\nu}_{1'2'}}{\omega + \omega_{\nu} - i\delta}\Bigr].
\label{respspec}
\ee
Similarly to the ones for the one-fermion and two-fermion propagators (\ref{spgfspec},\ref{resppp}), it satisfies the general quantum field theory requirements of locality and unitarity with the residues composed of the properly normalized matrix elements of the transition densities:
\be
\rho^{\nu}_{12} = \langle 0|\psi^{\dagger}_2\psi_1|\nu \rangle. 
\ee
They describe the weights of the pure particle-hole configurations on top of the ground state $|0\rangle$ in the model (ideally, exact) excited states $|\nu\rangle$ of the (even-even) $N$-particle system. The corresponding poles are the excitation energies of this system $\omega_{\nu} = E_{\nu} - E_0$.

Obviously, the spectral representation of the propagators given by  Eqs. (\ref{spgfspec},\ref{resppp},\ref{respspec}) are model independent: they are valid regardless how the many-body states $|n\rangle, |m\rangle$, $|\nu\rangle$, $|\mu\rangle$ and $|\varkappa\rangle$ are modeled. The sums in Eqs. (\ref{spgfspec},\ref{resppp},\ref{respspec}) run over both the discrete and continual sectors of the excitation spectra, i.e. formally complete.

Summarising, because of their simple relations to the key observables, fermionic propagators are important characteristics of strongly-coupled many-fermion systems, in particular, of atomic nuclei.
In Ref. \cite{LitvinovaSchuck2019} we have investigated the propagators of Eqs. (\ref{spgf}) and (\ref{phresp}) by generating the equations of motion for them. In the present work we will focus on the particle-particle Green function of Eqs. (\ref{ppGF},\ref{resppp}) by considering its time evolution and investigating its potential of describing nuclear pairing properties. 

\section{Equation of motion for the two-time pair correlation function}
\label{EOM}
\subsection{The model-independent EOM}

The time evolution of the correlation function of a fermionic pair can be investigated by the differentiation with respect to the time variable applied to Eq. (\ref{ppGFmod}):
\bea
\partial_t G_{12,1'2'}(t-t') = -i\delta(t-t')\langle [\psi_1\psi_2,{\psi^{\dagger}}_{2'}{\psi^{\dagger}}_{1'}]\rangle + \nonumber \\
+ \langle T[H,\psi_1\psi_2](t)({\psi^{\dagger}}_{2'}{\psi^{\dagger}}_{1'})(t')\rangle, \nonumber\\ 
\label{dtG}                           
\eea
where we defined
\be
[H,A](t) = e^{iHt}[H,A]e^{-iHt}
\ee
for an arbitrary operator $A$ and adopted the notation $G_{12,1'2'}(t-t') = G(12,1'2')$ for the two-time particle-particle propagator with  $t_1 = t_2 = t, t_{1'} = t_{2'} = t'$.
The commutators can be computed straightforwardly:
\bea
 [\psi_1\psi_2,{\psi^{\dagger}}_{2'}\psi^{\dagger}_{1'}] = \delta_{22'}\psi_1\psi^{\dagger}_{1'}  - \delta_{11'}\psi^{\dagger}_{2'}\psi_{2} -  \nonumber \\
 - \delta_{12'}\psi_{2}\psi^{\dagger}_{1'} + \delta_{21'}\psi^{\dagger}_{2'}\psi_{1},
\label{commnorm}
\eea
\be
[H,\psi_1\psi_2] = -(\varepsilon_1 + \varepsilon_2)\psi_1\psi_2 + [V,\psi_1\psi_2],\\
\ee
so that the first EOM takes the form:
\bea
(i\partial_t - \varepsilon_1 &-& \varepsilon_2)G_{12,1'2'}(t-t') = \delta(t-t'){\cal N}_{121'2'} + \nonumber \\
&+&i\langle  T[V,\psi_1\psi_2](t)({\psi^{\dagger}}_{2'}{\psi^{\dagger}}_{1'})(t')\rangle,
\label{EOM1}
\eea
where we introduced the norm matrix in the pp-channel as the ground state average of the commutator of Eq. (\ref{commnorm}):
\be
{\cal N}_{121'2'} = \langle [\psi_1\psi_2,{\psi^{\dagger}}_{2'}\psi^{\dagger}_{1'}] \rangle.
\ee
In the basis diagonalizing the one-body density matrix $\rho_{ij} = \langle \psi^{\dagger}_j \psi_i \rangle = \delta_{ij}n_i$ the norm matrix reads:
\be
{\cal N}_{121'2'} = \delta_{121'2'}(1- n_1 - n_2) = \delta_{121'2'}n_{12} 
\ee
with the antisymmetrized Kronecker symbol $\delta_{121'2'} = \delta_{11'}\delta_{22'} - \delta_{21'}\delta_{12'}$ and $n_{12} = 1-n_1-n_2$. The antisymmetrized Kronecker symbol
and the norm matrix satisfy the obvious relationships, which will be useful in the following context:
\bea
{\delta}_{121'2'} = -{\delta}_{211'2'} = -{\delta}_{122'1'} = {\delta}_{212'1'} = {\delta}_{1'2'12} \nonumber\\
{\cal N}_{121'2'} = -{\cal N}_{211'2'} = -{\cal N}_{122'1'} = {\cal N}_{212'1'} = {\cal N}_{1'2'12} \nonumber\\
\frac{1}{2}\sum\limits_{34}\delta_{1234}\delta_{341'2'} = \delta_{121'2'}, \ \ \   
\frac{1}{2}\sum\limits_{34}{\cal N}_{1234}{\cal N}^{-1}_{341'2'} = \delta_{121'2'}\nonumber\\
\eea
with the inverse norm defined as
\be
{\cal N}^{-1}_{121'2'} = \frac{\delta_{121'2'}}{1-n_1-n_2} = \delta_{121'2'}n^{-1}_{12}.
\ee
At this stage it is convenient to generate the second EOM. This can be done by differentiating the last term on the right hand side of the first EOM (\ref{EOM1})  with respect to the second time argument $t'$. Setting $F_{121'2'}(t-t') = i\langle  T[V,\psi_1\psi_2](t)({\psi^{\dagger}}_{2'}{\psi^{\dagger}}_{1'})(t')\rangle$, we come to the following equation:
\bea
(-i\partial_{t'} - \varepsilon_{1'} - \varepsilon_{2'})F_{121'2'}(t-t') = \nonumber\\ 
= -\delta(t-t')\langle[[V,\psi_1\psi_2],\psi^{\dagger}_{2'}\psi^{\dagger}_{1'}]\rangle + \nonumber\\ 
+i\langle  T[V,\psi_1\psi_2](t)[V,{\psi^{\dagger}}_{2'}{\psi^{\dagger}}_{1'}](t')\rangle .\nonumber\\
\label{EOM2}
\eea
The equation for the spectral image of the two-fermion propagator can be then obtained by combining Eqs. (\ref{EOM1},\ref{EOM2}) and subsequent Fourier transformation to the energy domain.
We define the spectral image $G_{12,1'2'}(\omega)$ as:
\be
G_{12,1'2'}(t-t') = \int\limits_{-\infty}^{\infty} \frac{d\omega}{2\pi} e^{-i\omega(t-t')}G_{12,1'2'}(\omega)
\ee 
and, thus, obtain:
\bea
G_{12,1'2'}(\omega) &=& G^{(0)}_{12,1'2'}(\omega) + \nonumber \\
&+& \frac{1}{4}\sum\limits_{343'4'}G^{(0)}_{12,34}(\omega)T_{343'4'}(\omega)G^{(0)}_{3'4',1'2'}(\omega)\nonumber\\  
\label{GTmatrix}
\eea
with the free pp-propagator introduced as:
\be
G^{(0)}_{12,1'2'}(\omega) = \frac{{\cal N}_{121'2'}}{\omega - \varepsilon_1 - \varepsilon_2}
\label{ppuncor}
\ee
and the renormalized kernel 
\be
T_{121'2'}(\omega) =  \frac{1}{4}\sum\limits_{343'4'}{\cal N}^{-1}_{1234}\Bigl( T^{(0)}_{343'4'} + T^{(r)}_{343'4'}(\omega)\Bigr){\cal N}^{-1}_{3'4'1'2'},
\ee
where $T^{(0)}_{343'4'} $ and $T^{(r)}_{343'4'}(\omega)$ are the Fourier images of the two last terms on the right hand side of Eq. (\ref{EOM2}), i.e.
\bea
T^{(0)}_{121'2'}(t-t') &=&  -\delta(t-t') \langle  [[V,\psi_1\psi_2],{\psi^{\dagger}}_{2'}{\psi^{\dagger}}_{1'}]\rangle\nonumber\\
T^{(r)}_{121'2'}(t-t') &=&  i\langle  T[V,\psi_1\psi_2](t)[V,{\psi^{\dagger}}_{2'}{\psi^{\dagger}}_{1'}](t')\rangle, \nonumber\\
\label{Tstdyn}
\eea
so that we have explicitly isolated the static part $T^{(0)}$ of the interaction kernel $T$ from its dynamical part $T^{(r)}$.
In full analogy with the case of the particle-hole response \cite{LitvinovaSchuck2019}, Eq. (\ref{GTmatrix}) can be transformed to an equation of the Dyson type
\bea
G_{12,1'2'}(\omega) &=& G^{(0)}_{12,1'2'}(\omega) + \nonumber \\
&+& \frac{1}{4}\sum\limits_{343'4'}G^{(0)}_{12,34}(\omega){K}_{343'4'}(\omega)G_{3'4',1'2'}(\omega)\nonumber\\  
\label{GDyson}
\eea
by introducing the new kernel $K(\omega)$ which can be obtained from $T(\omega)$ by retaining only the terms irreducible with respect to the uncorrelated pp-propagator 
(\ref{ppuncor}): 
\bea
T_{121'2'}(\omega) &=& K_{121'2'}(\omega) + \nonumber \\
&+& \frac{1}{4}\sum\limits_{343'4'}K_{1234}(\omega)G^{(0)}_{34,3'4'}(\omega)T_{3'4'1'2'}(\omega)\nonumber\\  
\label{Tmatrix}
\eea
or $K(\omega) = T^{(irr)}(\omega)$. Obviously, the removal of the reducible contributions affects only the dynamical part of $T$.
Remarkably, as in the case of the particle-hole response, Eq. (\ref{GDyson}) has the form of the Dyson equation. Its interaction kernel contains a static and a one frequency dependent parts, in full analogy to the Dyson equation for the one-body propagator. In other words, the Bethe-Salpeter equation for the two-time two-body Green functions can be regarded as Dyson Bethe-Salpeter equation (Dyson-BSE) \cite{Schuck2019}.

In the next two subsections we consider the static and dynamical parts of the kernel $T$ defined in Eq.(\ref{Tstdyn}) in the explicit form. 

\subsection{The static kernel}
\begin{figure}
\begin{center}
\includegraphics[scale=0.52]{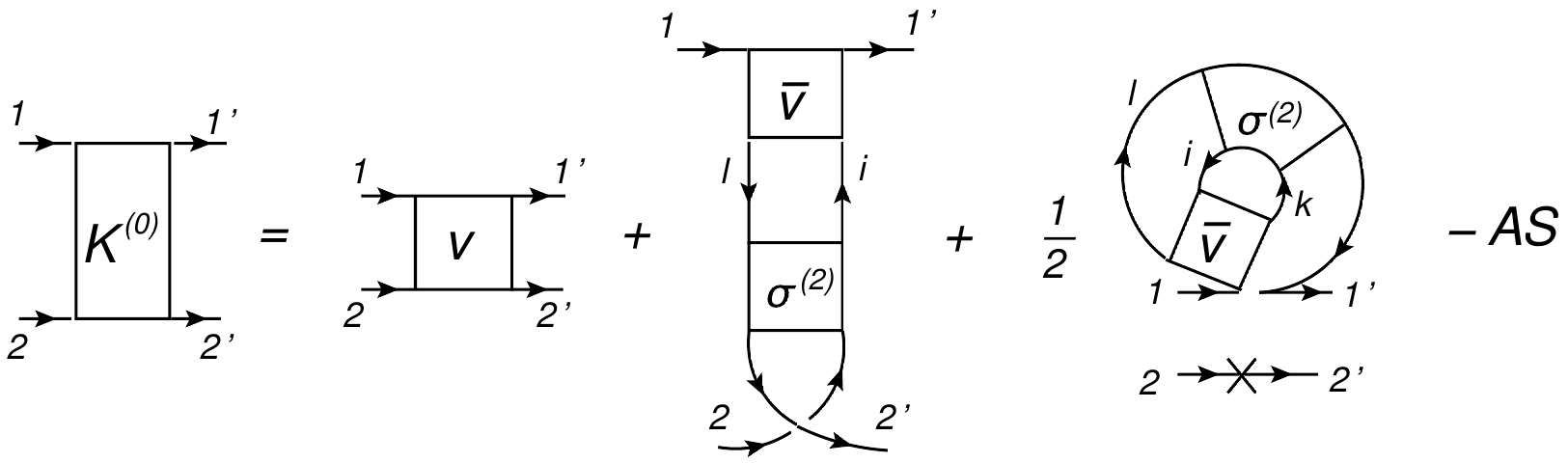}
\end{center}
\caption{Diagrammatic form of the static part ${K}^{(0)}$ of the interaction kernel given by Eq. (\ref{K0}). Lines with arrows represent one-fermion propagators. Rectangular blocks $v$ and $\bar v$ stand for the non-antisymmetrized and antisymmetrized bare two-fermion interaction and those marked with $\sigma^{(2)}$, together with the attached long fermionic lines,  for the fully correlated part of the two-body density. The abbreviation "AS" denotes the full antisymmetrization. The cross stands for the Kronecker delta symbol $\delta_{22'}$. Phases in front of each diagram are omitted.} 
\label{SEirrs}%
\end{figure}

The static part requires evaluating the commutator:
\bea
[V,\psi_1\psi_2] = \frac{1}{4}\sum\limits_{ijkl}{\bar v}_{ijkl}[{\psi^{\dagger}}_i{\psi^{\dagger}}_j\psi_l\psi_k,\psi_1\psi_2] = \nonumber\\
=\frac{1}{2}\sum\limits_{ijkl}{\bar v}_{ijkl}(\delta_{1j}\psi^{\dagger}_i\psi_{2} +  \delta_{2j}\psi_1\psi^{\dagger}_{i})\psi_l\psi_k = \nonumber\\
= \frac{1}{2}\sum\limits_{ijkl}{\bar v}_{ijkl}(\delta_{1j}\psi^{\dagger}_i\psi_{2} + \delta_{2j}\delta_{1i} - \delta_{2j}\psi^{\dagger}_{i}\psi_1)\psi_l\psi_k, \nonumber\\
\eea
so that
\bea
-T^{(0)}_{121'2'} &=& \langle[[V,\psi_1\psi_2],{\psi^{\dagger}}_{2'}{\psi^{\dagger}}_{1'}]\rangle = \frac{1}{2}\sum\limits_{kl}{\bar v}_{12kl}{\cal N}_{lk,1'2'} + \nonumber\\
&+&\langle\sum\limits_{ikl}{\bar v}_{i1kl}(\delta_{2'k}\psi^{\dagger}_i\psi_2\psi_l\psi^{\dagger}_{1'} + \delta_{1'k}\psi^{\dagger}_{2'}\psi^{\dagger}_{i}\psi_2\psi_l +\nonumber\\
&+&\frac{1}{2}\delta_{22'}\psi^{\dagger}_i\psi_l\psi_k\psi^{\dagger}_{1'} + \frac{1}{2}\delta_{21'}\psi^{\dagger}_{2'}\psi^{\dagger}_{i}\psi_l\psi_k)  - (1\leftrightarrow 2)\rangle.\nonumber\\
\label{T01}
\eea
By reorganizing the fermionic field operators into the two-body densities
\be
\rho_{ijkl} = \langle \psi^{\dagger}_{k}\psi^{\dagger}_{l}\psi_{j}\psi_{i}\rangle = \rho_{ik}\rho_{jl} - \rho_{il}\rho_{jk} + \sigma^{(2)}_{ijkl}
\ee
and introducing the mean-field single-particle energies 
\be
{\tilde\Sigma}_{11'} = \sum\limits_{l}{\bar v}_{1l1'l}n_l, \ \ \ \ \ \ \  {\tilde\Sigma}_{11'} = {\delta}_{11'}{\tilde\Sigma}_{1},
\ee
Eq. (\ref{T01}) can be written as follows:
\bea 
T^{(0)}_{121'2'} &=& \delta_{121'2'}n_{12}({\tilde\Sigma}_1 + {\tilde\Sigma}_2) + K^{(0)}_{121'2'}, \label{T02}\\
K^{(0)}_{121'2'} &=& {\bar v}_{121'2'}n_{12}n_{1'2'}  \nonumber \\
&-& \Bigl[ \Bigl( \sum\limits_{il}{\bar v}_{i12'l}\sigma^{(2)}_{l2i1'} + \frac{\delta_{22'}}{2}\sum\limits_{ikl} {\bar v}_{i1kl}\sigma^{(2)}_{kli1'}\Bigr)  \nonumber\\
&-& \Bigl(1'\leftrightarrow 2' \Bigr) \Bigr] -\Bigl[1\leftrightarrow 2 \Bigr], 
\label{K0}
\eea
which are consistent with the obvious antisymmetry properties of the static kernel: $T^{(0)}_{121'2'} = -T^{(0)}_{211'2'} = -T^{(0)}_{122'1'} = T^{(0)}_{212'1'}$. The first term on the right hand side of
Eq. (\ref{T02}) contains the mean-field single-particle energies which can be absorbed in the uncorrelated propagator, so that
\bea
G_{12,1'2'}(\omega) &=& {\tilde G}^{(0)}_{12,1'2'}(\omega) + \nonumber \\
&+& \frac{1}{4}\sum\limits_{343'4'}{\tilde G}^{(0)}_{12,34}(\omega){K}_{343'4'}(\omega)G_{3'4',1'2'}(\omega),\nonumber\\  
\label{GDyson1}
\eea
where
\be
{\tilde G}^{(0)}_{12,1'2'}(\omega) = \frac{{\cal N}_{121'2'}}{\omega - {\tilde\varepsilon}_1 - {\tilde\varepsilon}_2}, \ \ \ \ \ \ {\tilde\varepsilon}_1 = {\varepsilon}_1 + {\tilde\Sigma}_{1}
\label{ppuncortilde}
\ee
and the kernel $K$ does not contain the mean-field term in its static part while the dynamical part remains unchanged: ${K} = {\cal N}^{-1}({K}^{(0)} + K^{(r)}){\cal N}^{-1}$. The obtained static part of the interaction kernel $K^{(0)}_{121'2'}$ is shown in Fig. \ref{SEirrs} in the diagrammatic representation. The form of Eq. (\ref{K0}) for the static kernel is consistent with Ref. \cite{DukelskyRoepkeSchuck1998}, where it was derived in a similar way.  As one sees, the static kernel is composed, besides the bare interaction term, of an instantaneous exchange of a particle-hole ($ph$) correlation function between the two particles (often called screening) and a renormalization of the single-particle energies by the same $ph$ correlation function. In applications it is important to treat both contributions on the same footing, since often quite important cancellations between the two contributions occur.

In the simplest approximation, the EOM (\ref{GDyson1}) for the particle-particle propagator can be considered with only the static part $K^{(0)}$ of the kernel neglecting completely the contribution from $K^{(r)}$. In analogy to the case of the particle-hole response function, such an approach is equivalent to the self-consistent particle-particle random phase approximation. 
In the description of superfluid systems, both the particle-hole and particle-particle response functions are coupled in the framework of the self-consistent quasiparticle random phase approximation, or SCQRPA, which demonstrates great success in applications to the two-level pairing model  \cite{Rabhi2002}.  The part $K^{(r)}$ of the kernel is associated with dynamical processes induced by the medium, which produce an interplay of screening and antiscreening effects on the pairing gaps \cite{Cao2006,Guo2019a}.

\subsection{The dynamical kernel}
The dynamical part of the interaction kernel can be calculated with the aid of the commutator:
\bea
[V,\psi^{\dagger}_{2'}\psi^{\dagger}_{1'}] = \frac{1}{4}\sum\limits_{ijkl}{\bar v}_{ijkl}[{\psi^{\dagger}}_i{\psi^{\dagger}}_j\psi_l\psi_k,\psi^{\dagger}_{2'}\psi^{\dagger}_{1'}] = \nonumber\\
=\frac{1}{2}\sum\limits_{ijkl}\psi^{\dagger}_{i}\psi^{\dagger}_{j}(\delta_{k2'}\psi_{l}\psi^{\dagger}_{1'} +  \delta_{k1'}\psi^{\dagger}_{2'}\psi_l){\bar v}_{ijkl} = \nonumber\\
=\frac{1}{2}\sum\limits_{ijkl}\psi^{\dagger}_{i}\psi^{\dagger}_{j}(\delta_{k2'}\delta_{l1'} - \delta_{k2'}\psi^{\dagger}_{1'}\psi_{l} +  \delta_{k1'}\psi^{\dagger}_{2'}\psi_l){\bar v}_{ijkl},\nonumber \\
\eea
so that
\bea
T^{(r)}_{121'2'}(t-t') = \frac{i}{4}\sum\limits_{ijkl}\sum\limits_{mnpq}{\bar v}_{ijkl}\langle T\bigl[(\delta_{1j}\psi^{\dagger}_i\psi_{2} +  \delta_{2j}\psi_1\psi^{\dagger}_{i})\times\nonumber\\
\times\psi_l\psi_k\bigr](t)\bigl[\psi^{\dagger}_{m}\psi^{\dagger}_{n}(\delta_{p2'}\psi_{q}\psi^{\dagger}_{1'} +  \delta_{p1'}\psi^{\dagger}_{2'}\psi_q)\bigr](t')\rangle{\bar v}_{mnpq}\nonumber\\
\eea
or
\bea
T^{(r)}_{121'2'}(t-t') = \frac{i}{4}\times\nonumber\\
\times\sum\limits_{ikl}\sum\limits_{mnq}\Bigl[{\bar v}_{i1kl}\langle T(\psi^{\dagger}_i\psi_2\psi_l\psi_k)(t)(\psi^{\dagger}_m\psi^{\dagger}_n\psi^{\dagger}_{2'}\psi_q)(t')\rangle {\bar v}_{mn1'q} +  \nonumber\\
+ {\bar v}_{i1kl}\langle T(\psi^{\dagger}_i\psi_2\psi_l\psi_k)(t)(\psi^{\dagger}_m\psi^{\dagger}_n\psi_q\psi^{\dagger}_{1'})(t')\rangle {\bar v}_{mn2'q}
+ \nonumber \\
+ {\bar v}_{i2kl}\langle T(\psi_1\psi^{\dagger}_i\psi_l\psi_k)(t)(\psi^{\dagger}_m\psi^{\dagger}_n\psi^{\dagger}_{2'}\psi_q)(t')\rangle {\bar v}_{mn1'q} +\nonumber\\
+ {\bar v}_{i2kl}\langle T(\psi_1\psi^{\dagger}_i\psi_l\psi_k)(t)(\psi^{\dagger}_m\psi^{\dagger}_n\psi_q\psi^{\dagger}_{1'})(t')\rangle {\bar v}_{mn2'q}\Bigr].  \nonumber\\
\label{Tr}
\eea
\begin{figure}
\begin{center}
\includegraphics[scale=0.52]{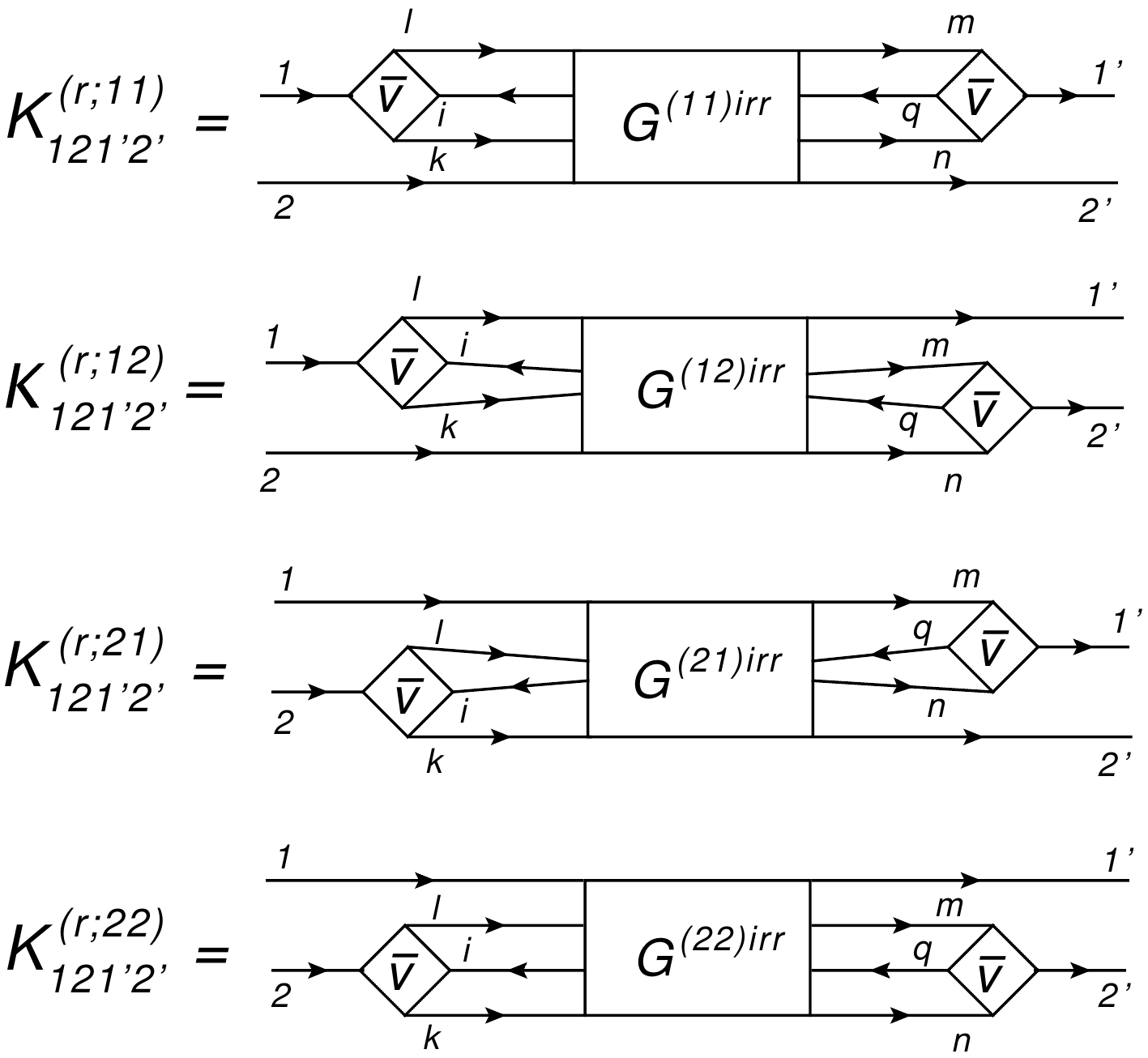}
\end{center}
\caption{Diagrammatic representation of the dynamical part $K^{(r)}(\omega) = T^{(r)irr}(\omega)$ of the interaction kernel. The rectangular blocks $G^{(ab)irr}$, together with the long fermionic lines, denote the four-fermion correlation functions irreducible in the particle-particle channel. Factor $i/4$ in front of each diagram is omitted.}
\label{SEirrd}%
\end{figure}

Thus, in complete analogy to the case of the particle-hole response \cite{LitvinovaSchuck2019}, the dynamical kernel of the EOM for the particle-particle propagator is determined by the 
irreducible four-fermion correlation functions. The nature of these correlation functions is, however, different. Namely, each term of Eq. (\ref{Tr}) contains a propagator of three particles and one hole ($3p-1h$). Therefore, an approximate cluster expansion truncated on the two-body level should contain all possible products of the particle-particle and particle-hole correlation functions. For instance, the internal propagator in the first term of Eq. (\ref{Tr}) can be factorized as follows:
\bea
&G&^{(11)irr}_{2lkq,2'nmi}(t-t') = \langle T(\psi^{\dagger}_i\psi_2\psi_l\psi_k)(t)(\psi^{\dagger}_m\psi^{\dagger}_n\psi^{\dagger}_{2'}\psi_q)(t')\rangle^{irr} \nonumber \\
&\approx& \langle T(\psi^{\dagger}_i\psi_2)(t)(\psi^{\dagger}_{2'}\psi_q)(t')\rangle \langle T(\psi_l\psi_k)(t)(\psi^{\dagger}_m\psi^{\dagger}_n)(t')\rangle \nonumber \\
&+& \langle T(\psi^{\dagger}_i\psi_k)(t)(\psi^{\dagger}_{n}\psi_q)(t')\rangle \langle T(\psi_l\psi_2)(t)(\psi^{\dagger}_m\psi^{\dagger}_{2'})(t')\rangle  \nonumber\\
&+& \langle T(\psi^{\dagger}_i\psi_l)(t)(\psi^{\dagger}_{m}\psi_q)(t')\rangle \langle T(\psi_k\psi_2)(t)(\psi^{\dagger}_n\psi^{\dagger}_{2'})(t')\rangle \nonumber\\
&-& {\cal AS},
\label{G4_11_0}
\eea
if the fully correlated three-body and four-body terms are neglected. Using the definitions (\ref{ppGFmod},\ref{phresp}), the $3p-1h$ propagator of Eq. (\ref{G4_11_0}) can be rewritten as
\bea
&G&^{(11)irr}_{2lkq,2'nmi}(t-t') \approx -R_{i2,q2'}(t-t')G_{lk,nm}(t-t')  \nonumber \\ 
&-& R_{ik,qn}(t-t')G_{l2,2'm}(t-t') - R_{il,qm}(t-t')G_{k2,2'n}(t-t') \nonumber \\ &-& {\cal AS}. \label{G4_11} 
\eea
The other three $3p-1h$ correlation functions can be factorized similarly, so that 
\bea
&G&^{(12)irr}_{2lkq,1'nmi}(t-t') \approx R_{i2,qn}(t-t')G_{lk,m1'}(t-t')  \nonumber \\ 
&+& R_{il,q1'}(t-t')G_{k2,nm}(t-t') + R_{ik,qm}(t-t')G_{l2,n1'}(t-t') \nonumber \\ &-& {\cal AS}  \label{G4_12} 
\eea
\bea
&G&^{(21)irr}_{1lkq,2'nmi}(t-t') \approx R_{i1,qm}(t-t')G_{kl,n2'}(t-t')  \nonumber \\ 
&+& R_{ik,q2'}(t-t')G_{l1,mn}(t-t') + R_{il,qn}(t-t')G_{1k,2'm}(t-t') \nonumber \\ &-& {\cal AS}  \label{G4_21} 
\eea
\bea
&G&^{(22)irr}_{1lkq,1'nmi}(t-t') \approx -R_{i1,q1'}(t-t')G_{lk,nm}(t-t')  \nonumber \\ 
&-& R_{il,qm}(t-t')G_{1k,n1'}(t-t') - R_{ik,qn}(t-t')G_{1l,m1'}(t-t') \nonumber \\ &-& {\cal AS}, \label{G4_22} 
\eea
with ${\cal AS}$ being the remaining antisymmetrizations, and the irreducible kernel takes the form:
\bea
K^{(r)}_{121'2'}(t-t') = \frac{i}{4}\times\nonumber \\
\times\sum\limits_{ikl}\sum\limits_{mnq}
\Bigl[{\bar v}_{i1kl}G^{(11)irr}_{2lkq,2'nmi}(t-t') {\bar v}_{mn1'q} + \nonumber\\
+ {\bar v}_{i1kl}G^{(12)irr}_{2lkq,1'nmi}(t-t'){\bar v}_{mn2'q}
+ \nonumber \\
+ {\bar v}_{i2kl}G^{(21)irr}_{1lkq,2'nmi}(t-t') {\bar v}_{mn1'q} +\nonumber \\
+ {\bar v}_{i2kl}G^{(22)irr}_{1lkq,1'nmi}(t-t') {\bar v}_{mn2'q}\Bigr] =  \nonumber\\
= K^{(r;11)}_{121'2'}(t-t') + K^{(r;12)}_{121'2'}(t-t') + \nonumber \\
+ K^{(r;21)}_{121'2'}(t-t') + K^{(r;22)}_{121'2'}(t-t').
\label{Kr}
\eea

The cluster expansion of Eq. (\ref{G4_11_0}), thereby, shows how the many-body problem can be truncated on the level of two-body correlations in the pairing, or particle-particle, channel. 
The presence of the particle-hole propagators in the dynamical kernel of the pair propagator expresses explicitly the coupling of the particle-hole and particle-particle channels  in the many-body systems. This points to the formulation of the truncated EOM's for fermionic correlation functions in a closed form  \cite{SchuckTohyama2016a,Olevano2018,LitvinovaSchuck2019}. 

Alternatively to the symmetric form of the dynamical kernel $K^{(r)}$, which is obvious from Eq. (\ref{Kr}) and Fig. \ref{SEirrd}, it can be expressed via the three-fermion propagator, if only the first EOM (\ref{EOM1}) is generated. 
More generally,  the EOM for the $n$-fermion Green function in the non-symmetric form contains $(n-1)$ and $(n+1)$-fermion Green functions, thus, generating an infinite hierarchy of coupled equations.
This hierarchy can be truncated at any level, and truncation on the $n$-body level would mean that the dynamical kernels of the EOM's for the $n$-body and higher propagators are approximated by the cluster expansions that involve up to $n$-fermion correlation functions.

As far as the present example and the pairing correlations are concerned, one may notice that the theory truncated on the two-body level does not involve the anomalous one-fermion Green functions of the Gorkov's type \cite{Gorkov1958}. Instead, the pairing correlations influence the one-fermion propagator  via the two-fermion Green functions of the particle-particle type, which enter the dynamical kernel of the one-body equation of motion \cite{LitvinovaSchuck2019}. This fact has been noted already in Ref. \cite{Terasaki2002} and briefly discussed there in a different context.  On one hand, the approach with the two-fermion particle-particle correlation functions may look more complicated because it requires  a solution of the coupled one-fermion and two-fermion EOM's. But, on the other hand, it avoids working in the doubled quasiparticle space, that can be technically quite demanding in the models with explicit dynamical kernels. Besides that, the present approach overcomes problems related to the particle number violation, which are inherent in the BCS and Bogoliubov's theories.

\section{Emergent phonons and mapping to the particle-phonon coupling}
\label{PVC}

The EOM for the fermionic pair propagator (\ref{GDyson}) in the energy (frequency) domain contains the Fourier transform of the dynamical kernel (\ref{Kr}). As all the terms of its propagators' expansion (\ref{G4_11}-\ref{G4_22}) consist of non-contracted products of one particle-particle and one particle-hole propagators, they can be treated with the aid of the following generic transformation: 
\bea
[R_{12,1'2'}G_{34,3'4'}](\omega) = \int\limits_{-\infty}^{\infty}d\tau e^{i\omega\tau} R_{12,1'2'}(\tau)G_{34,3'4'}(\tau)  \nonumber\\
= -i\Bigl[\sum\limits_{\nu\mu}\frac{ \rho_{21}^{\nu}\rho_{2'1'}^{\nu\ast}\alpha_{43}^{\mu}\alpha_{4'3'}^{\mu\ast} }{\omega - \omega_{\nu} - \omega_{\mu}^{(++)} + i\delta} 
- \sum\limits_{\nu\varkappa}\frac{ \rho_{12}^{\nu\ast}\rho_{1'2'}^{\nu}\beta_{34}^{\varkappa\ast}\beta_{3'4'}^{\varkappa} }{\omega + \omega_{\nu} + \omega_{\varkappa}^{(--)} - i\delta}\Bigr].\nonumber\\
\eea
Then, each product should be contracted with two matrix elements of the two-fermion interaction $\bar v$, as given by Eq. (\ref{Kr}), so that the components of the dynamical kernel take the form:
\bea
K^{(r;11)}_{121'2'}(\omega) = -\frac{i}{4}\sum\limits_{ikl}\sum\limits_{mnq}{\bar v}_{i1kl}\times\nonumber\\
\times \Bigl([R_{i2,q2'}G_{lk,nm}](\omega) + [R_{ik,qn}G_{l2,2'm}](\omega) + \nonumber \\ + [R_{il,qm}G_{k2,2'n}](\omega) 
-{\cal AS}\Bigr) {\bar v}_{mn1'q},
\label{K11}
\eea
\bea
K^{(r;12)}_{121'2'}(\omega) = \frac{i}{4}\sum\limits_{ikl}\sum\limits_{mnq}{\bar v}_{i1kl}\times\nonumber\\
\times\Bigl([R_{i2,qn}G_{lk,m1'}](\omega) + [R_{il,q1'}G_{k2,nm}](\omega) +  \nonumber \\ + [R_{ik,qm}G_{l2,n1'}](\omega) 
-{\cal AS}\Bigr) {\bar v}_{mn2'q},
\label{K12}
\eea
\be
K^{(r;21)}_{121'2'}(\omega) = K^{(r;12)}_{212'1'}(\omega),
\label{K21}
\ee
\be
K^{(r;22)}_{121'2'}(\omega) = K^{(r;11)}_{212'1'}(\omega).
\label{K22}
\ee
\begin{figure},
\begin{center}
\includegraphics[scale=0.50]{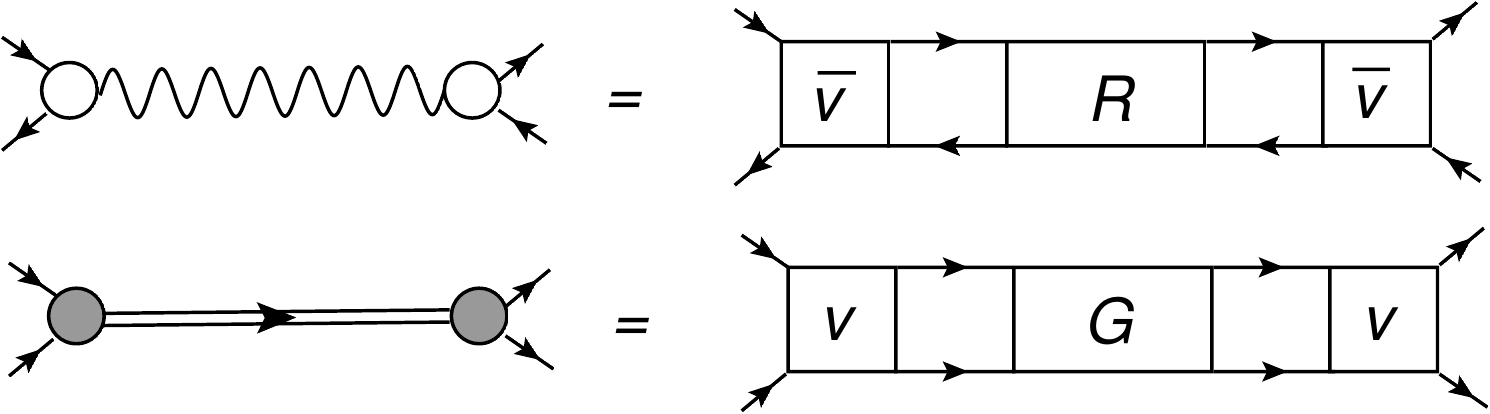}
\end{center}
\caption{The exact mapping of the phonon vertices (circles) and propagators (wavy lines and double lines) onto the bare interaction and the particle-hole ($R$) and particle-particle ($G$) correlation functions. Top: normal (particle-hole) phonon, bottom: pairing (particle-particle) phonon,  as introduced in Eqs. (\ref{mappingph},\ref{mappingpp}), respectively.}
\label{PVCmap}%
\end{figure}
The second and the third terms in Eq. (\ref{K11}) are already symmetric, but we have retained both of them to keep the analogy with Eq. (\ref{K12}). As one can see in the following, these two terms form the contribution of the self-energy type with the normal phonon, that is expressed by the first line of Eq. (\ref{K11phon}).    
Similarly to Ref. \cite{LitvinovaSchuck2019}, one can introduce the vertices $g^{\nu(\pm)}$ and propagators $D_{\nu}^{(\pm)}(\omega)$ of the normal phonons:
\bea
g^{\nu(\sigma)}_{13} = \delta_{\sigma,+}g^{\nu}_{13} + \delta_{\sigma,-}g^{\nu\ast}_{31}, \ \ \ \ 
g^{\nu}_{13} = \sum\limits_{24}{\bar v}_{1234}\rho^{\nu}_{42}, \nonumber \\
D_{\nu}^{(\sigma)}(\omega) = \frac{\sigma}{\omega - \sigma(\omega_{\nu} - i\delta)}, \ \ \ \
\label{gDPVCph}
\eea 
as well as the particle-phonon coupling amplitude 
\bea
\Gamma^{(ph)}_{13',1'3} =  \sum\limits_{242'4'}{\bar v}_{1234}R^{(ph)}_{24,2'4'}(\omega){\bar v}_{4'3'2'1'} = \nonumber \\ = 
\sum\limits_{\nu,\sigma=\pm} g^{{\nu}(\sigma)}_{13}D^{(\sigma)}_{\nu}(\omega)g^{\nu(\sigma)\ast}_{1'3'}.
\label{mappingph}
\eea
Analogously,  the vertices $\gamma^{\mu(\pm)}$ and propagators $\Delta^{(\pm)}_{\mu}(\omega)$ of the pairing, or superfluid, phonons are 
\bea
\gamma^{\mu(+)}_{12} = \sum\limits_{34} v_{1234}\alpha_{34}^{\mu}, \ \ \ \ \ \ \gamma_{12}^{\varkappa(-)} = \sum\limits_{34}\beta_{34}^{\varkappa}v_{3412} \nonumber \\ 
\Delta^{(\sigma)}_{\mu}(\omega) = \frac{\sigma}{\omega - \sigma(\omega_{\mu}^{(\sigma\sigma)} - i\delta)}
\eea
with the corresponding coupling amplitude $\Gamma^{(pp)}_{12,1'2'}(\omega) $:
\bea
\Gamma^{(pp)}_{12,1'2'}(\omega) = \sum\limits_{343'4'}{v}_{1234}G^{(pp)}_{43,3'4'}(\omega){v}_{4'3'2'1'} = \nonumber \\
= \sum\limits_{\mu,\sigma=\pm1} \gamma^{\mu(\sigma)}_{12}\Delta^{(\sigma)}_{\mu}(\omega)\gamma^{\mu(\sigma)\ast}_{1'2'},
\label{mappingpp}
\eea
where the index $\mu$ runs over both addition and removal modes.
The mapping to emergent particle-hole and particle-particle (pairing) phonons is depicted in Fig. \ref{PVCmap} in the diagrammatic form.  
\begin{figure*}
\begin{center}
\includegraphics[scale=0.85]{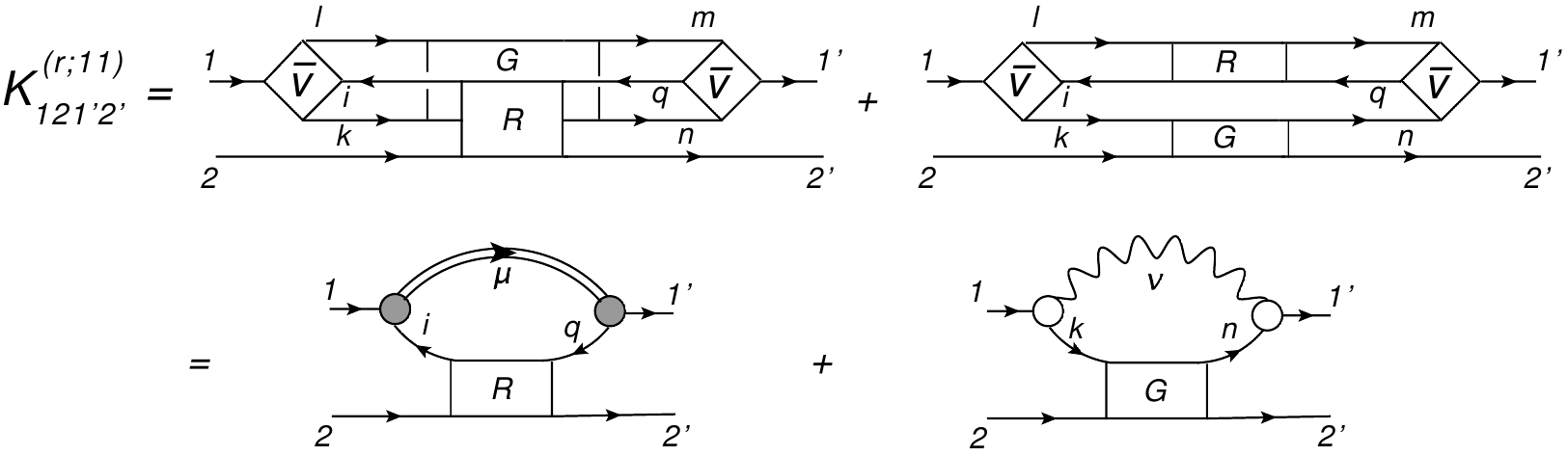}
\end{center}
\caption{The $K^{(r;11)}$ component of the dynamical interaction kernel. The diagrammatic conventions of Fig. \ref{PVCmap} are employed.}
\label{K11_dyn}%
\end{figure*}
\begin{figure*}
\begin{center}
\includegraphics[scale=0.85]{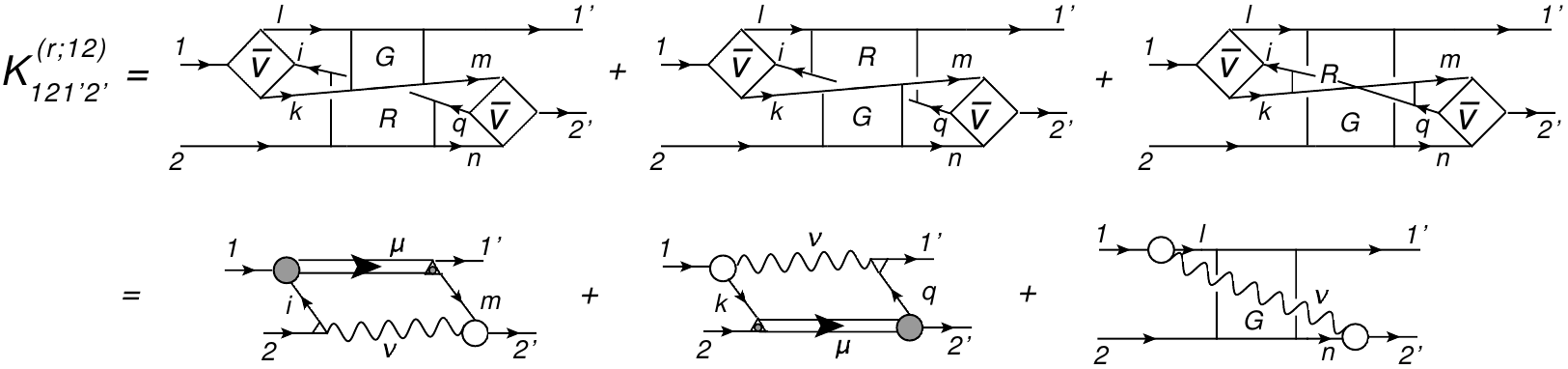}
\end{center}
\caption{The $K^{(r;12)}$ component of the dynamical interaction kernel. The same diagrammatic conventions as in the previous figures are used. The additional little empty and filled triangles stand for the normal $\rho^{\nu}$ and pairing $\alpha^{\mu}$ transition densities, respectively.}
\label{K12_dyn}%
\end{figure*}

With the notions of these emergent phonons, the first component of the dynamical kernel associated with the self-energy graph (the upper line of Fig. \ref{SEirrd}) takes the following form:  
\bea
&\ &K^{(r;11)}_{121'2'}(\omega) = \nonumber \\
&=&\Bigl[ 
\sum\limits_{kn;\nu\mu}\frac{g_{1k}^{\nu}\alpha_{2k}^{\mu}\alpha_{2'n}^{\mu\ast} g_{1'n}^{\nu\ast}}{\omega - \omega_{\nu} - \omega_{\mu}^{(++)} + i\delta} -%
\sum\limits_{kn;\nu\varkappa}\frac{g_{k1}^{\nu\ast}\beta_{k2}^{\varkappa\ast}\beta_{n2'}^{\varkappa} g_{n1'}^{\nu}}{\omega + \omega_{\nu} + \omega_{\varkappa}^{(--)} - i\delta} \Bigr]
\nonumber\\
&+&\Bigl[ \sum\limits_{iq;\nu\mu}\frac{\gamma_{1i}^{\mu(+)}\rho_{2i}^{\nu}\rho_{2'q}^{\nu\ast} \gamma_{1'q}^{\mu(+)\ast}}{\omega - \omega_{\nu} - \omega_{\mu}^{(++)} + i\delta} 
- \sum\limits_{iq;\nu\varkappa}\frac{\gamma_{i1}^{\varkappa(-)\ast}\rho_{i2}^{\nu\ast}\rho_{q2'}^{\nu} \gamma_{q1'}^{\varkappa(-)}}{\omega + \omega_{\nu} + \omega_{\varkappa}^{(--)} - i\delta}\Bigr]  \nonumber\\
\label{K11phon}
\eea
while the second component represented by the "twisted" graph in the second line of Fig. \ref{SEirrd} reads:
\bea
&\ &K^{(r;12)}_{121'2'}(\omega) = 
\nonumber \\
&=&-\Bigl[ \sum\limits_{im;\nu\mu}\frac{\gamma_{1i}^{\mu(+)}\rho_{2i}^{\nu}\alpha_{1'm}^{\mu\ast} g_{2'm}^{\nu\ast}}{\omega - \omega_{\nu} - \omega_{\mu}^{(++)} + i\delta} 
- \sum\limits_{im;\nu\varkappa}\frac{\gamma_{i1}^{\varkappa(-)\ast}\rho_{i2}^{\nu\ast}\beta_{m1'}^{\varkappa} g_{m2'}^{\nu}}{\omega + \omega_{\nu} + \omega_{\varkappa}^{(--)} - i\delta}\Bigr]
\nonumber\\ 
&-&
\Bigl[ \sum\limits_{kq;\nu\mu}\frac{g_{1k}^{\nu}\alpha_{2k}^{\mu} \rho_{1'q}^{\nu\ast}\gamma_{2'q}^{\mu(+)\ast}}{\omega - \omega_{\nu} - \omega_{\mu}^{(++)} + i\delta} 
- \sum\limits_{kq;\nu\varkappa}\frac{g_{k1}^{\nu\ast}\beta_{k2}^{\varkappa\ast} \rho_{q1'}^{\nu}\gamma_{q2'}^{\varkappa(-)}}{\omega + \omega_{\nu} + \omega_{\varkappa}^{(--)} - i\delta}\Bigr]
\nonumber\\
&-& 
\Bigl[ 
\sum\limits_{ln;\nu\mu}\frac{g_{1l}^{\nu}\alpha_{2l}^{\mu}\alpha_{1'n}^{\mu\ast} g_{2'n}^{\nu\ast}}{\omega - \omega_{\nu} - \omega_{\mu}^{(++)} + i\delta} -%
\sum\limits_{ln;\nu\varkappa}\frac{g_{l1}^{\nu\ast}\beta_{l2}^{\varkappa\ast}\beta_{n1'}^{\varkappa} g_{n2'}^{\nu}}{\omega + \omega_{\nu} + \omega_{\varkappa}^{(--)} - i\delta} \Bigr].
\nonumber\\
\label{K12phon}
\eea
 The two remaining components $K^{(r;21)}_{121'2'}(\omega)$ and $K^{(r;22)}_{121'2'}(\omega)$ can be found from Eqs. (\ref{K11phon},\ref{K12phon}) with the help of the symmetry relations of Eqs. (\ref{K21},\ref{K22}).  
 
 The diagrammatic representation of the components $K^{(r;11)}_{121'2'}(\omega)$ and $K^{(r;12)}_{121'2'}(\omega)$ of the dynamical kernel are shown in Figs. \ref{K11_dyn} and \ref{K12_dyn}, respectively. In the case of $K^{(r;11)}_{121'2'}(\omega)$, the mapping to the PVC leads to two topologically similar terms of the self-energy type with the pairing and normal phonons.   
The "twisted" component $K^{(r;12)}_{121'2'}(\omega)$ contains a typical phonon-exchange term (the third term in Eq. (\ref{K12phon}) and in Fig. \ref{K12_dyn}) with the normal phonon, however, its counterpart with the single pairing phonon would violate the particle number conservation and, therefore, is absent in this component. Instead, mixed contributions of the normal and pairing phonons appear, as it is clear from the graphical form of the first two terms of Eq. (\ref{K12phon}) in Fig. \ref{K12_dyn}.

\section{A static limit and the pairing gap equation} 
\label{Gap}

\begin{figure*}
\begin{center}
\includegraphics[scale=0.50]{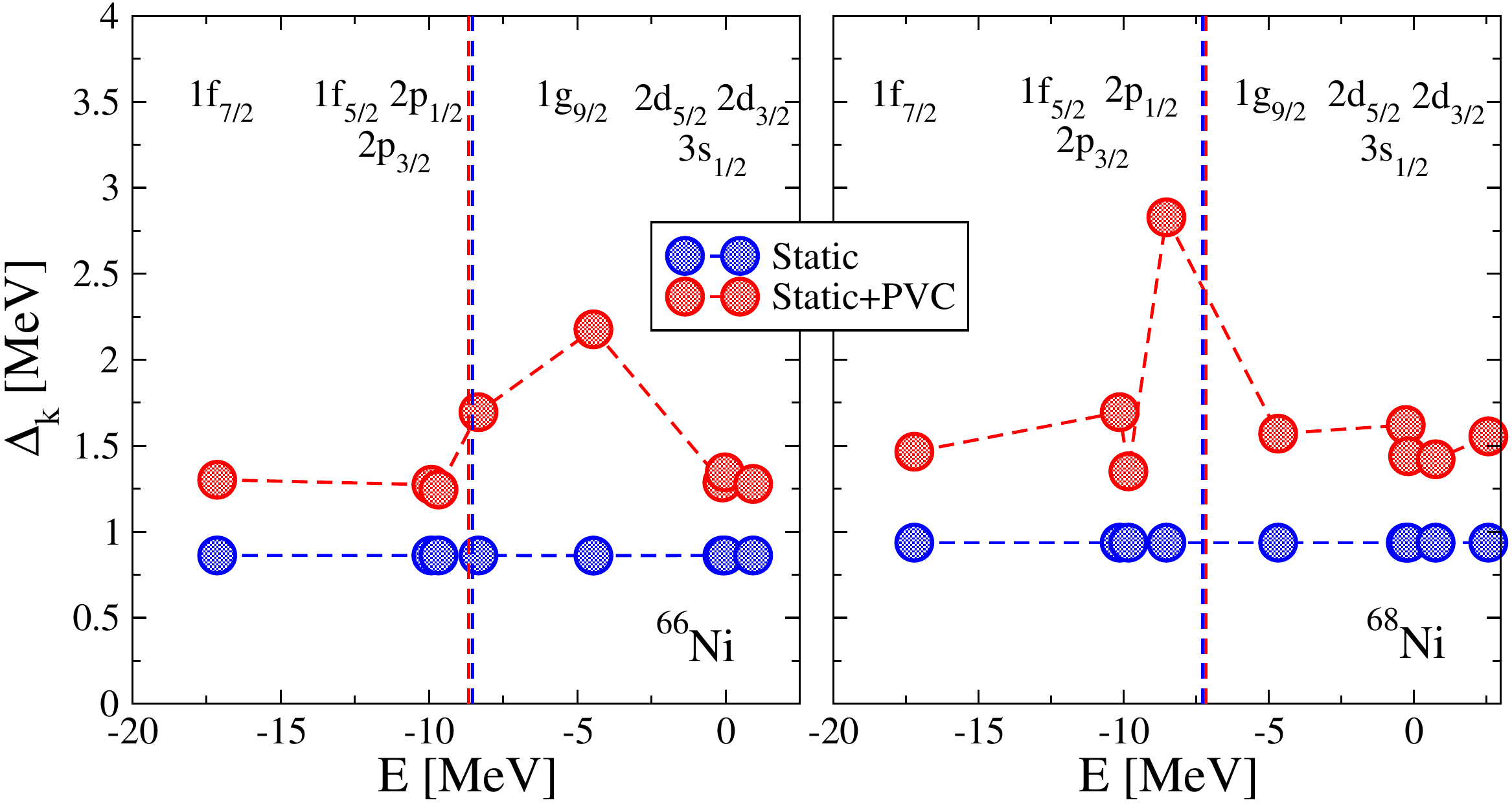}
\end{center}
\caption{The state-dependent pairing gaps in $^{66}$Ni and $^{68}$Ni calculated in the purely static constant-gap approximation with a phenomenological kernel (blue symbols) and taking into account the dynamical PVC effects  (red symbols). The vertical dashed lines mark the Fermi energy.}
\label{ni-gaps}%
\end{figure*}

From Eq. (\ref{GDyson}) it is easy to obtain the equations for the pairing transition densities  $\alpha^{\mu}$ and $\beta^{\varkappa}$. Indeed, considering the frequency argument of $G_{12,1'2'}(\omega)$
in the vicinity of a pole in the $(N+2)$-particle system $\omega = \omega_s$, leads to the following equation for $\alpha^{\mu}$:
\be
\alpha^s_{21} = \frac{1-n_1 - n_2}{\omega_s - {\tilde\varepsilon}_1 - {\tilde\varepsilon}_2} \frac{1}{4}\sum\limits_{343'4'}\delta_{1234}K_{343'4'}(\omega_s)\alpha^s_{4'3'}
\label{alpha}
\ee
and a similar equation for $\beta^{\varkappa}$.
Furthermore, if the ground state of the reference nucleus is approximated by the BCS-like approach, where \cite{ZV.book}
\be
n_1 = v_1^2 = \frac{E_1-({\tilde\varepsilon}_1-\lambda)}{2E_1}, \ \ \ \ \ \ \ E_1 = \sqrt{({\tilde\varepsilon}_1-\lambda)^2+\Delta_1^2 }
\ee
and $\lambda$ being the chemical potential, the pairing gap $\Delta_1$ can be related to the pairing transition density as 
\be
\Delta_1 = 2E_1\alpha^{s}_{{\bar 1}1}, 
\ee
and at $\omega_s = 2\lambda$ Eq. (\ref{alpha}) takes the form of the well-known pairing gap equation:
\be
\Delta_1 = -\sum\limits_{2}{\cal V}_{1{\bar 1}2{\bar 2}}\frac{\Delta_2}{2E_2},
\label{gap}
\ee
where the bar denotes the conjugate or the time-reversed state \cite{RingSchuck1980} and the interaction matrix elements read:
\be
{\cal V}_{121'2'} = \frac{1}{4}\sum\limits_{34}\delta_{1234}K_{341'2'}(2\lambda) = \frac{1}{2}\Bigl(K^{(0)}_{121'2'} + K^{(r)}_{121'2'}(2\lambda) \Bigr).
\label{Kgap}
\ee 
The integral part of the gap equation (\ref{gap}), thus, contains all the microscopic effects of the kernel $K$ "on shell", regardless the approximations made for its static $K^{(0)}$ and dynamical $K^{(r)}$ parts.  



\section{Calculations: details, results and discussion}
\label{Results}

\begin{figure*}
\begin{center}
\includegraphics[scale=0.50]{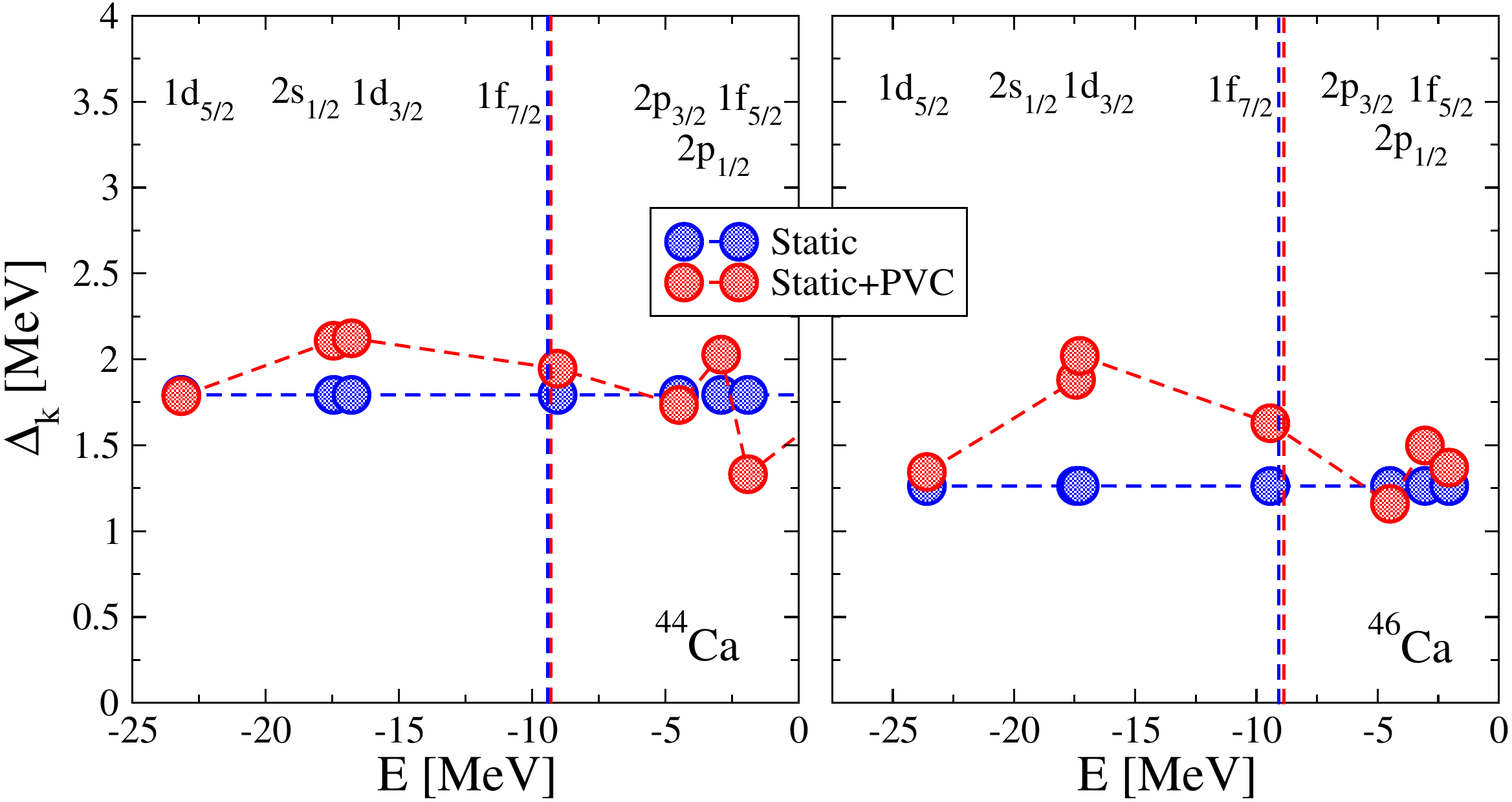}
\end{center}
\caption{Same as in Fig. \ref{ni-gaps} but for $^{44}$Ca and $^{46}$Ca.}
\label{ca-gaps}%
\end{figure*}

In order to test the developed approach in realistic conditions, we have performed some illustrative calculations for finite nuclei.  As in Ref. \cite{LitvinovaSchuck2019}, in these calculations we were focused mainly on the dynamical kernel, but now for the propagator of the fermion pair in the static limit, that is determined by the nuclear pairing gaps. Thus, we solved Eq. (\ref{gap}) with the kernel of Eq. (\ref{Kgap}) that includes both the static and dynamical parts, together with the usual BCS constraint on the average particle number \cite{50BCS}. In this first study we approximated the static part of the interaction by the phenomenological 'monopole force' which is detailed, for instance,  in Ref. \cite{LitvinovaRingTselyaev2008}. Consistently, we used the basis of the relativistic mean field (RMF) with NL3 parametrization \cite{Lalazissis1997} to approximate the one-body part of the Hamiltonian.  

In this first study of the PVC effects on the pairing gaps in a relativistic framework, the dynamical kernel $K^{(r)}(\omega)$ was computed in the leading approximation that (i) omits the exchange of the pairing phonons, thus, keeping only the last terms shown in Figs. \ref{K11_dyn}, \ref{K12_dyn} and (ii) neglects the particle-particle correlations in the 'G' parts of those terms. This form of the dynamical kernel corresponds to the leading 'two-quasiparticle plus phonon' ($2q\otimes$phonon) approximation, which is commonly employed in the nuclear field theories.  Although more sophisticated approaches are already available for the particle-hole response from Refs. \cite{LitvinovaRingTselyaev2013,LitvinovaSchuck2019,Robin2019}, here we investigated only the major PVC effect on the nuclear pairing gaps.  The latter are known to be linked to the observed odd-even mass differences as discussed, for instance, in Ref. \cite{Bender2000}, and can  thus be extracted from the nuclear mass tables  \cite{Audi2002}.
\begin{figure*}
\begin{center}
\includegraphics[scale=0.50]{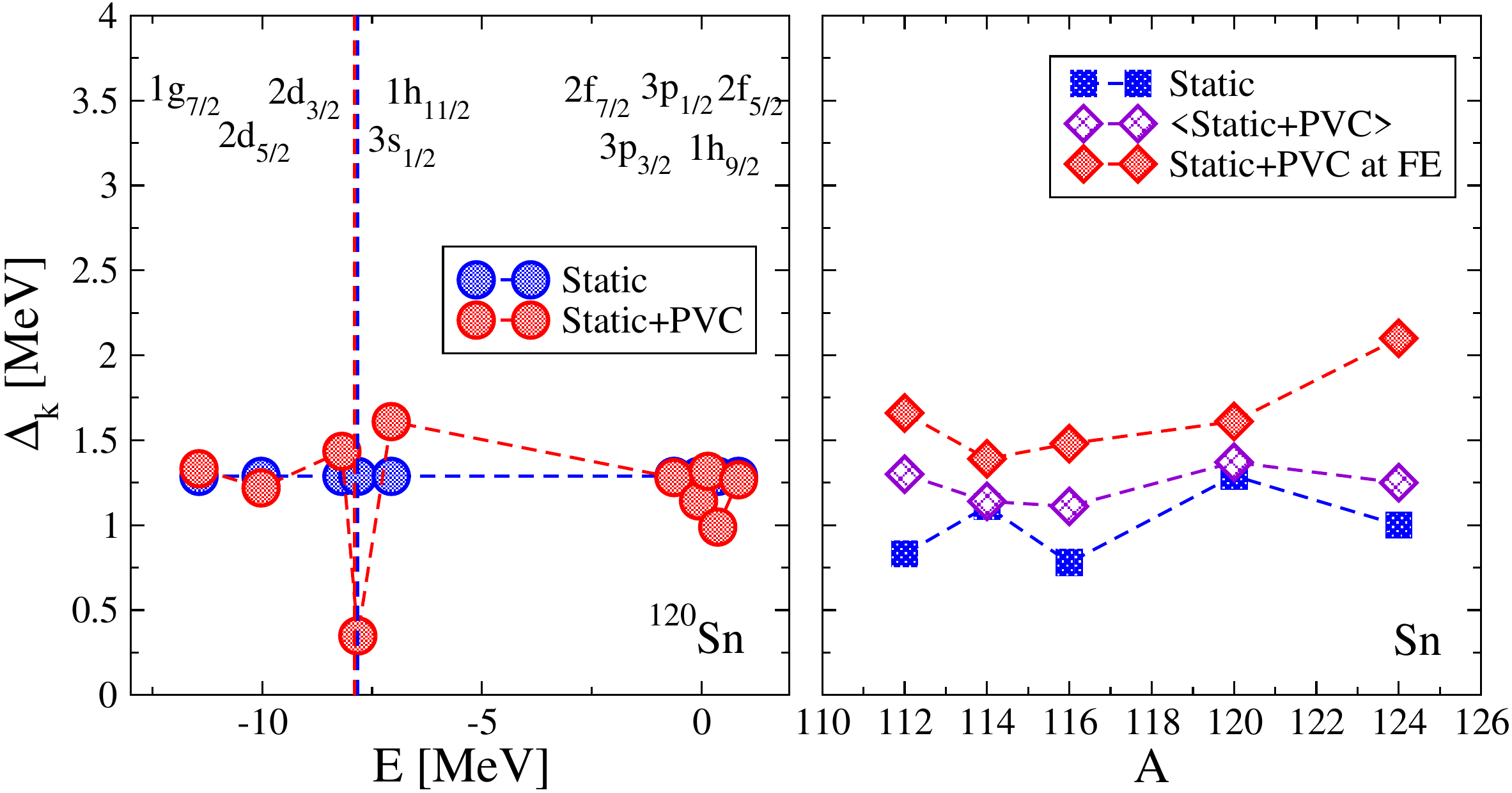}
\end{center}
\caption{Pairing gaps in tin isotopes. Left panel: The state-dependent pairing gaps in $^{120}$Sn without (blue circles) and with (red circles) the PVC effects. Right panel: The average pairing gaps in stable tin isotopes calculated without (blue squares) and with (violet diamonds) the PVC effects, compared to the peak values of the pairing gaps around the Fermi energy obtained in the 'Static+PVC' calculations (red diamonds). }
\label{sn-gaps}%
\end{figure*}

Since we do not address an ab-initio calculation of the pairing gaps, but rather keep the static part of the interaction kernel purely phenomenological, the calculations have, besides the six RMF parameters, one free parameter which is adjusted to reproduce  the average pairing gap in a $\approx$ 20 MeV energy window around the Fermi energy, or chemical  potential. In this way, it is possible to determine the relative contributions of the static and dynamical kernels.
Based on the RMF for the given nucleus, the phonon vertices $g^{\nu}$ and their frequencies $\omega_{\nu}$ were extracted from the relativistic quasiparticle random phase approximation (RQRPA) \cite{PaarRingNiksicEtAl2003} calculations  with the empirical pairing gap (see below) . The latter were performed for the angular momenta and parities $J_{\nu}^{\pi_{\nu}}$ = 2$^+$, 3$^-$, 4$^+$, 5$^-$, 6$^+$ forming the phonon model space commonly used in the PVC applications. This model space was slightly truncated to select the modes which provide the most important contributions. In particular, in this work we used the same truncation criteria as in Ref. \cite{LitvinovaSchuck2019}. After that, Eq.  (\ref{gap}) with the kernel of Eq. (\ref{Kgap}) and the BCS particle number constraint were solved in a self-consistent cycle, where the value of the state-dependent pairing gaps and the chemical potential were determined with a 10$^{-3}$ MeV accuracy. 

Fig. \ref{ni-gaps} illustrates the results of calculations of the neutron pairing gaps in the two open-shell nickel isotopes $^{66}$Ni and $^{68}$Ni. First we performed the calculation with both the static phenomenological and the PVC kernels, where the strength parameter in front of the static kernel is tuned in a way that the resulting averaged pairing gaps reproduce the empirical value extracted from Ref. \cite{Audi2002} with the help of the three-point formula \cite{Bender2000}. The results of these calculations are shown by the red symbols and marked as 'Static+PVC' on the legends. After that, the PVC, or dynamical, part of the interaction kernel was turned off and the calculations were repeated with the same static kernel. These results are shown by the blue symbols and marked as 'Static'. Thus, we isolated the PVC effects of the dynamical kernel $K^{(r)}$ that can be assessed quantitatively by the difference between the two results. 

The first observation from Fig. \ref{ni-gaps} is that the calculations with the PVC contribution produce larger average pairing gaps. The latter contribution is a result of an interplay between the self-energy terms of the type shown in Fig. \ref{K11_dyn}, their exchange counterparts shown in Fig.  \ref{K12_dyn} as well as  the remaining $K^{(r;22)}$ and $K^{(r;21)}$ components with analogous structure.  As in the case of the dynamical kernel of the particle-hole response \cite{LitvinovaRingTselyaev2007,LitvinovaRingTselyaev2008}, the exchange contributions contain phase factors varying from state to state, while the self-energy terms add up coherently. This results in a partial cancellation of the self-energy and exchange contributions, which is known, in particular, to moderate the effect of broadening of the giant resonances caused by the PVC mechanism. This kind of cancellation is also known to be more pronounced in the monopole channel \cite{LitvinovaRingTselyaev2007}, which is the case for the singlet pairing that we consider here. Nevertheless, the net average pairing gaps obtained with the PVC for $^{66}$Ni and $^{68}$Ni are considerably larger than corresponding gaps of the purely static character. 
Another observation is that the PVC mechanism brings a sizable state dependence of the pairing gap as compared to the calculations with nearly constant static kernel. In all the cases investigated in the present work, which include nickel, calcium and tin isotopic chains, we found a remarkable state dependence of the neutron pairing gaps obtained with the PVC kernel. In particular, an enhancement of the gaps is found for the single-particle states at or near the Fermi surface.  Our calculations show 153\% ($1g_{9/2}$) and 200\% ($2p_{1/2}$) enhancement of the peak values of the neutron pairing gap in these nickel isotopes, as compared to their 'static' values  while the average pairing gap increases by factors 1.77 and 1.69, respectively.

Fig. \ref{ca-gaps} shows the results of similar calculations for two calcium isotopes, $^{44}$Ca and $^{46}$Ca. In these two open-shell nuclei as well as in $^{42}$Ca which we also investigated, the enhancement of the average  pairing gap due to the PVC effects is less pronounced than in nickel isotopes adding up to the factors of 1.06 and 1.25, respectively. This may indicate a stronger cancellation between the self-energy and exchange PVC terms in these nuclei.  However, one can still notice a remarkable state dependence of $\Delta_k$.  In both $^{44}$Ca and $^{46}$Ca it peaks at the  $2s_{1/2}$ and $1d_{3/2}$ hole states located next to the Fermi surface. The peak values of the neutron pairing gaps exhibit about 20\% and 60\% enhancement with respect to their 'static' values in the two calcium isotopes, respectively. 

In heavier systems the trends are similar. The left panel of Fig. \ref{sn-gaps} illustrates the behavior of the pairing gaps in $^{120}$Sn. Again, we find that the PVC contribution  to the interaction kernel brings a noticeable state dependence, especially at the Fermi surface. The average value, however, remains nearly unchanged. In the right panel of Fig. \ref{sn-gaps} we plot the average pairing gaps in some stable tin isotopes obtained with and without the PVC effects.  One can see that the PVC enhancement of the average value varies and can reach 50 \% as it occurs, for instance, in $^{112}$Sn. The peak values of the pairing gaps at the Fermi surface obtained in the 'Static+PVC' calculation are also plotted and show a soft minimum in the middle of the shell.

While calculations for all nuclei under study show maxima for the pairing gap value around the Fermi energy due to the PVC, for the deep hole states the effect of the PVC varies from nucleus to nucleus. In some systems like $^{66,68}$Ni and $^{112,116}$Sn the pairing gap values for the deeply bound states are noticeably enhanced by the PVC. In other systems, such as $^{44,46}$Ca and $^{114,120}$Sn, a relatively weak effect of the PVC on the pairing gap values for the deeply bound states is observed. We have found a correlation between this effect and the pairing gap enhancement at its peak value relative to the gap values for the deep hole states in the 'static + PVC' calculations: the more pronounced peaks are associated with the larger increments of the pairing gap for the deeply bound states above its 'static' values. In general, the resulting values of the pairing gap are determined by a complex interplay of the PVC matrix elements and the shell structure around the Fermi energy. However, the particular choice of constraining the average pairing gap values by the experimental odd-even mass differences  may also play a role. For instance, one may choose to identify the pairing gap at the Fermi surface with the experimental pairing gap instead, and in this case the trends for the deeply bound states may also vary. Consistent calculations based on a bare nucleon-nucleon interaction are needed for more definite conclusions.

The behavior of the pairing gaps around the Fermi surface is in a qualitative agreement with Refs. \cite{BarrancoBrogliaGoriEtAl1999,Terasaki2002}
and may occur due to the specific structure of the PVC contributions to the interaction kernel. Nevertheless, our attempts to perform calculations with only the dynamical PVC kernel did not result in the realistic pairing gaps even for the states around the Fermi energy, in contrast to the latter two studies. The calculations of Refs. \cite{BarrancoBrogliaGoriEtAl1999,50BCS} employed the nuclear field theory with the effective particle-phonon Hamiltonian and the Bloch-Horowitz second-order perturbation theory for the pairing interaction induced by the particle-phonon coupling, while in Refs. \cite{Avdeenkov1999,Terasaki2002} a generalized Dyson equation was solved for the single-quasiparticle propagator in the doubled quasiparticle space, which treats pairing and PVC effects on the equal footing.  The latter method is still based on the Gorkov factorization and on the use of the anomalous one-fermion Green functions. As a consequence, these approaches commonly contain the dynamical kernel with only coherent PVC contributions, that may explain the overall stronger PVC effects. 

The more recent study of Ref. \cite{Idini2012} is based on the formalism of Ref. \cite{VanderSluys1993}. The latter postulates the excitation operator in the particle-phonon coupling form, which is equivalent to introducing the interaction of Eq. (\ref{mappingph}) between the nucleons, in addition to the mean field, and generates the equation of motion for this operator.  This strategy allows for a consistent  calculation scheme for both the single-quasiparticle strength distributions and the state-dependent pairing gaps. In Ref.  \cite{Idini2012}, the Argonne $v_{14}$ and V$_{low k}$ potentials were employed for the static part of the nucleon-nucleon interaction, Skyrme SLy4 parametrization for the mean field and phenomenological separable interaction for determining the PVC coupling vertices and phonon frequencies. In this hybrid approach, both the state-dependent pairing gaps and the single-quasiparticle strength distributions in tin isotopes were investigated. The PVC effects on the pairing gaps were stipulated by the explicit phonon-exchange term and the so-called Z-factors, which are analogous to the $K^{(r;12)}$ and $K^{(r;11)}$ parts of our dynamical kernel, respectively. The numerical results are also quantitatively closer to those obtained in the present work, namely the dynamical effects caused by the PVC induces additional $\sim 20-40\%$ contributions to the pairing gaps, as compared to the calculations without PVC.  A comparison with Ref. \cite{Lombardo2001}, which investigated the role of the self-energy contributions responsible for the reduction of the quasiparticle strength at the Fermi surface in neutron matter, is also in agreement with our findings: overall, the self-energy contributions introduce the reduction while the phonon-exchange terms produce the screening or antiscreening of the pairing gaps.

The present implementation is considered as only the initial step towards a consistent microscopic theory of nuclear superfluidity beyond the standard BCS approximation. After quantifying the contribution of the dynamical kernel $K^{(r)}$ to the observed pairing gaps, the next natural move would be considering an accurate calculation of the static kernel $K^{(0)}$.  Instead of employing the simple monopole-force ansatz, the static kernel should be computed based on more realistic effective or bare interactions, that can use the insights from both the relativistic \cite{KucharekRing1991,SerraRummelRing2001,Serra2001,TianMaRing2009,Afanasjev2015a,50BCS} and non-relativistic \cite{Idini2012,Signoracci2015,Soma2014a,50BCS} studies. 
%
Finding a convenient auxiliary mean field and the associated basis is crucial for such calculations. Many groups use the harmonic oscillator basis, some employ the basis of the Woods-Saxon potential. A mean field of the contemporary energy density functionals can be chosen as well. After generating the mean-field basis, the RPA calculations with the corresponding effective interaction can be performed for the response functions in various channels. The response functions and two-body densities extracted from these calculations can serve as ingredients for initiating the iterative procedure. The procedure would start with the calculation of both static and dynamical kernels using the bare interaction contracted with the phenomenological two-body correlation functions, where the latter include both the response functions and the two-body densities. The outcome of the first as well as of the subsequent iterations are the single-fermion and the two-fermion correlation functions. In the second iteration, these correlation functions should replace the phenomenological ones used in the dynamical kernels in the first iteration and, thus, the procedure of recycling the one-body and two-body propagators can be continued until the convergence is achieved. Fast RPA solvers, for instance, the finite-amplitude method (FAM), can be engaged in such calculations to facilitate them after the FAM is upgraded with the dynamical kernels.   
Adding the knowledge about the treatment of two-fermion propagators with particles in the continuum, as outlined in Refs. \cite{Kamerdzhiev1998,KhanSandulescuGrassoEtAl2002}, would be also very instructive for applications to loosely bound and light nuclei.

For the nuclear matter and neutron matter, one of the main objectives at the moment is to get screening in the different spin-isospin channels under control. For instance, in the recent Ref. \cite{Guo2019a} calculations for the triplet pairing gap with both the static and dynamical kernels in the BCS equation have been performed. The calculations were based on the G-matrix interaction obtained from the Argonne AV18 potential and the meson-exchange three-body force. The G-matrix was used as the static kernel while RPA solutions obtained in the Landau limit with the same G-matrix were employed for calculating the dynamical kernel, or the induced interaction. The self-energy modification due to the medium polarization was taken into account via the Z-factors. Thus, as compared to the present paper, the calculations of Ref. \cite{Guo2019a} were done in a more advanced manner for the static kernel, but somewhat less accurate for the dynamical one. The obtained results allowed for important conclusions regarding the competition between the singlet and triplet pairing, where the latter typically occurs at high densities and the former dominates in the low-density regime. 

In Ref. \cite{Urban2020} finite-temperature calculations of the singlet pairing gap in dilute neutron matter were performed. The authors investigated the pairing gaps and the critical temperature of the superfluid phase transition using the $V_{low k}$ interaction derived from AV18 for the static kernel of the pairing gap equation and Skyrme interaction for the RPA vertices entering the dynamical kernel. They showed, in particular, that at higher densities the full RPA leads to stronger screening than the Landau approximation. It was noted, in general, that the pairing gap and the phase transition temperature are sensitive to the approximation used to describe the medium polarization effects. Therefore, in the long term, more complete and consistent calculations are desirable also for nuclear and neutron matters.
%

\section{Summary and outlook}
\label{Summary}

We introduce a many-body approach to the pairing correlation function in fermionic systems. The equation of motion method is formulated for the two-time two-fermion propagator in the particle-particle channel in a strongly-coupled medium. The EOM for this propagator takes the form of the Dyson Bethe-Salpeter equation, where both the static and dynamical interaction kernels are derived from the underlying bare two-fermion interaction. The exact symmetric dynamical kernel, which contains a four-fermion propagator, is approximated by a cluster decomposition into the two-fermion propagators of both particle-particle and particle-hole type.  In this way, the nuclear many-body problem is truncated at the level of two-body correlation functions whose EOM's, together with those for the one-fermion particle-hole correlation function discussed in Ref. \cite{LitvinovaSchuck2019} form a closed system of integral equations.

Although a complete solution of such a system is not yet available for finite nuclei, some aspects of the formulated approach can be studied for these systems. For instance, the resulting particle-particle correlation function appears to be related to the observables associated with the nuclear superfluidity. The equation for the pairing gap, which is directly related to a residue of the two-time particle-particle propagator, is therefore formulated as a static limit of the EOM for this propagator. Assuming the ground state  wave function of the BCS type, a BCS-like equation for the pairing gap is obtained. The interaction kernel of this equation, as the one of the corresponding EOM, has the purely static part as well as the dynamical part taken in the static limit. The latter contribution thus represents an extension of the BCS approximation to the inclusion of higher complexity correlations. 

We investigated the effects of this additional term on pairing gaps in medium-light and medium-heavy nuclei. Namely, the neutron pairing gaps were calculated for calcium, nickel and tin isotopes.  The developed method was implemented numerically on the base of quantum hadrodynamics and relativistic mean field. The beyond-mean-field effects on the pairing gaps are found quite pronounced. They lead to a sizable  state dependence of the pairing gaps with the tendency to an enhancement around the Fermi surface, in a qualitative agreement with existing NFT calculations. We found, however, that the static part of the interaction gives a relatively large contribution to the pairing gap values. Thus, we conclude  that this part should be also accurately determined from the underlying microscopic interaction. This is recognized as the most natural further advancement that will be addressed by future effort.

\section*{Acknowledgements}
This work is supported by the US-NSF Career Grant PHY-1654379.
%

\bibliography{Bibliography_Sep2019}
\end{document}